\documentclass[sn-mathphys,Numbered]{sn-jnl}

\usepackage{graphicx}%
\usepackage{multirow}%
\usepackage{amsmath,amssymb,amsfonts}%
\usepackage{amsthm}%
\usepackage{mathrsfs}%
\usepackage[title]{appendix}%
\usepackage{xcolor}%
\usepackage{textcomp}%
\usepackage{manyfoot}%
\usepackage{booktabs}%
\usepackage{algorithm}%
\usepackage{algorithmicx}%
\usepackage{algpseudocode}%
\usepackage{listings}%
\usepackage{adjustbox}
\usepackage{mathdots}

\theoremstyle{thmstyleone}%
\theoremstyle{thmstyletwo}%

\theoremstyle{thmstylethree}%

\begin{document}

\title[Article Title]{
  \begin{center}
    \textbf{Space-time refraction\\ of space-time wave packets}
  \end{center}
}

\author*{\fnm{Zeki} \sur{Hayran}}\email{z.hayran@imperial.ac.uk}

\author*{\fnm{John B.} \sur{Pendry}}\email{j.pendry@imperial.ac.uk}

\affil{The Blackett Laboratory, Department of Physics, Imperial College London, London SW7 2AZ, United Kingdom}

\abstract{Space-time modulation of refractive index can produce synthetically moving interfaces with arbitrary apparent velocities, including superluminal motion, offering new ways to control light in dynamic media. On the other hand, space-time wave packets are structured waves whose spatio-temporal spectra lie on tilted space-time planes, so their group velocity can be programmed, including superluminal values, even in a uniform medium. Here we develop a general theory of space-time refraction for such structured waves at a planar moving interface and show how a single boundary reshapes their velocity content. By identifying the invariants of a translating boundary, we obtain refraction laws for baseband, X--wave, and sideband packets that apply for arbitrary interface velocities and connect smoothly to static and purely temporal limits. These laws reveal regimes of ``space-time anomalous optical push broom,'' where a moving interface compresses a wide range of incident velocities into a narrow transmitted band, and ``velocity spectral optical fission,'' where an incident X--wave splits into two propagation-invariant branches with distinct velocities. The combined freedom to prepare waves with superluminal group velocity and to prescribe equally unconstrained interface speeds points toward reconfigurable time gating, optical buffering, velocity multiplexing, and controlled emission in moving media, and provides a route to photonic settings capable of emulating dynamical effects traditionally associated with gravitational or quantum processes.}

\keywords{space-time refraction, space-time wave packets, moving refractive index fronts, group velocity control, X--waves, space-time metamaterials, optical horizons, analogue gravity}

\maketitle

\section{Introduction}\label{sec1}

Controlling the flow of electromagnetic waves is a central theme in optics and photonics. The traditional approach relies on spatial structuring, from refractive components to photonic crystals and metasurfaces \cite{zheludev2012metamaterials,simovski2020introduction,butt2021recent}. In parallel, \textit{purely time-modulated media} have emerged as a powerful complementary tool \cite{galiffi2022photonics}, where uniform-in-space but time-varying refractive index changes enable frequency conversion \cite{hayran2021spectral, tirole2024second}, temporal reflection \cite{moussa2023observation, jones2024time}, time-reversal \cite{vezzoli2018optical}, and routes to overcome conventional limits \cite{hayran2023using, hayran2021capturing, hayran2024beyond, ciabattoni2025observation}. When spatial and temporal variations act together, \textit{space-time modulation} unlocks qualitatively new behaviour, including broadband and quantum-emission processes \cite{belgiorno2010quantum, sloan2022controlling, pendry2024qed}, nonreciprocal transport \cite{yu2009complete}, and analogue Fresnel-drag effects \cite{huidobro2019fresnel}, as well as related phenomena \cite{engheta2023four}. A key development in this direction is the concept of \textit{synthetic motion}, in which space-time modulation is engineered so that the refractive index profile drifts across space with a prescribed apparent velocity \cite{gaafar2019front, harwood2025space}. By shaping the driving wave, this synthetic boundary can be tuned from deeply subluminal to effectively superluminal \cite{gaafar2019front, harwood2025space}, providing a flexible and experimentally accessible way to emulate moving interfaces in optical materials.

A parallel viewpoint shifts the focus from engineering the medium to engineering the electromagnetic wave itself. Space-time wave packets (STWPs) are pulsed beams that propagate in the \(z\)-direction without diffraction or dispersion. In a uniform nondispersive medium, each plane-wave component lies on the light cone \(k_{x}^{2}+k_{z}^{2}=(n\omega/c)^{2}\). An STWP is obtained by selecting only those components that lie on the intersection of this cone with a tilted spectral plane. The slope of that plane in the \((k_{z},\omega)\) projection fixes a single axial group velocity \(v_{\mathrm g}= \mathrm d\omega/\mathrm dk_{z}=c\tan\theta\), where \(\theta\) is the spectral tilt defined in Fig. \ref{intro} \cite{yessenov2021refraction, yessenov2022space}. This freedom makes it possible to launch families of packets that share the same carrier frequency and spatial profile but approach the interface with widely different group velocities. Together with the independent choice of the interface velocity \cite{harwood2025space}, this provides separate control over the two key velocities that govern the interaction, the motion of the wave and the motion of the boundary, and makes STWPs natural probes of synthetic motion.

The interaction between such packets and moving boundaries has seen only limited theoretical attention. A recent inspirational theoretical analysis demonstrated that a moving refractive index front can strongly reshape a normally incident baseband packet, altering its group velocity and spectral form \cite{wang2025spatiotemporal}. While this study highlighted the opportunities offered by moving boundaries for controlling structured fields, it focused on a single wave packet family. Other families include sideband packets with shifted spectral content and X--wave packets that maintain a localized, propagation-invariant profile. These distinct spectral structures suggest qualitatively different refraction behaviour \cite{yessenov2021refraction}, however their transformation across a moving interface and the influence of oblique incidence remain unexplored to our knowledge.

Here we address this gap and develop a general theory of space-time refraction for baseband, sideband, and X--wave packets at arbitrary incidence. We identify the invariants that govern their transformation, obtain the corresponding refraction laws, and show how these laws connect continuously to both spatial- and temporal-interface limits, with full derivations and limiting cases presented in Supplementary Section S1. The resulting framework clarifies the distinct behaviour of different packet families, reveals singular conditions associated with optical horizons \cite{horsley2023quantum}, and provides a foundation for using STWPs to study dynamic media. To illustrate the range of possibilities, we present two example applications, ``space-time optical push broom'', in which a moving boundary compresses a wide spread of incident velocities into a narrow transmitted band, and ``velocity spectral optical fission'', in which an incident X--wave splits into two branches with distinct velocities, highlighting the diverse forms of velocity and spectral control enabled by a moving boundary.

\section{Main}

\begin{figure}
 \centering
 \begin{adjustbox}{width=1\linewidth, center}
  \includegraphics[keepaspectratio]{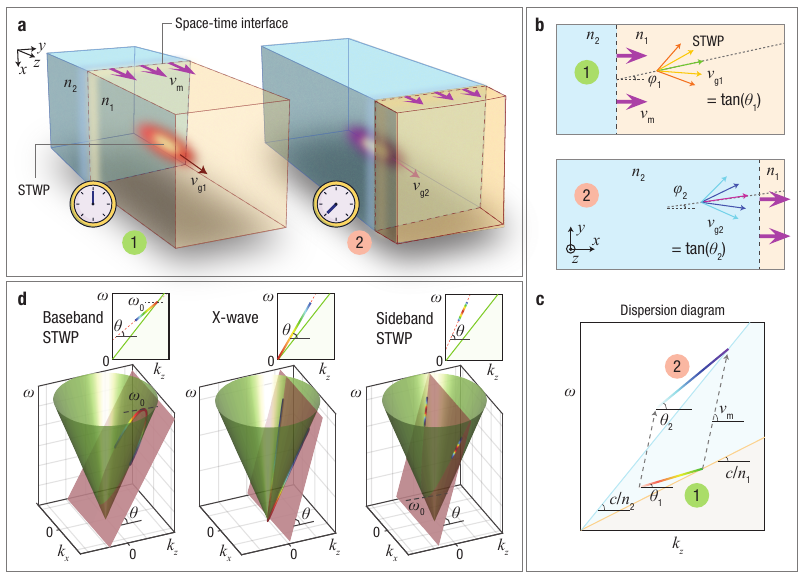}
  \end{adjustbox}
\caption{\textbf{Interaction of STWPs with a moving refractive index interface.}
\textbf{a}, Three dimensional schematic of a STWP encountering a planar space-time (ST) interface that separates two homogeneous media with refractive indices \(n_{1}\) and \(n_{2}\). Labels \textbf{1} and \textbf{2} mark the packet before and after interacting with the moving interface, which travels along \(z\) with velocity \(v_{\mathrm m}\). The corresponding group velocities are \(v_{\mathrm g1}\) and \(v_{\mathrm g2}\).
\textbf{b}, Two dimensional view of the same configuration. The STWP axis forms an angle \(\phi_{j}\) with the interface normal, while its spectrum is characterised by the tilt angle \(\theta_{j}\) in the \((k_{z},\omega)\) plane, with \(v_{\mathrm g j}= c\tan\theta_{j}\). The spectral tilt \(\theta_{j}\) determines the group velocity and is distinct from the incidence angle \(\phi_{j}\).
\textbf{c}, Space-time refraction in the \((k_{z},\omega)\) dispersion diagram at \(k_{x}=0\) for normal incidence. The red and blue lines denote the light lines for indices \(n_{1}\) and \(n_{2}\), and the refraction plane imposed by the moving interface maps an incident spectral line with tilt \(\theta_{1}\) to a transmitted line with tilt \(\theta_{2}\).
\textbf{d}, Spectral construction of the three STWP families considered in this work. Top row, intersections of the spectral planes with the \((k_{z},\omega)\) section, showing baseband STWPs, X--waves, and sideband STWPs; green lines indicate the corresponding light lines. Bottom row, full three dimensional \((k_{x},k_{z},\omega)\) representation, where green cones denote the light cones and coloured curves show the spectral loci that define each STWP family.}

    \label{intro}
\end{figure}

We begin by specifying the geometry of the moving interface and the STWPs to which the refraction laws apply. Figure \ref{intro} summarises the configuration: panels \ref{intro}(a) and \ref{intro}(b) show the real-space arrangement before and after the packet encounters the interface, panel \ref{intro}(c) shows the corresponding construction in the \((k_{z},\omega)\) dispersion diagram, and panel \ref{intro}(d) visualises the spectral support of the three STWP families in full \((k_{x},k_{z},\omega)\) space.

We consider a planar interface separating two homogeneous, nondispersive dielectrics with refractive indices \(n_{1}\) and \(n_{2}\). The interface is uniform along its tangent direction \(x\) and moves rigidly along its normal \(z\) with velocity \(v_{\mathrm m}\), as sketched in Fig. \ref{intro}(a,b). In the laboratory frame a STWP in medium \(j\) is incident from the left and is converted into a transmitted STWP on the right (see Fig. \ref{intro}(c)). The packet axes form angles \(\phi_{1}\) and \(\phi_{2}\) with the interface normal, with \(\phi_{1}=\phi_{2}=0\) corresponding to normal incidence. Throughout we assume an impedance matched interface, either because \(\varepsilon\) and \(\mu\) are matched or because the transition is implemented adiabatically, so reflection is negligible and we focus on the transmitted packet.

The refraction process is governed by two exact invariants at the moving interface. Translational symmetry along \(x\) enforces conservation of the tangential component \(k_{x}\), while invariance of the co-moving normal coordinate \(z-v_{\mathrm m} t\) enforces conservation of the Doppler-type combination \cite{liberal2024spatiotemporal}
\begin{equation}
T = k_{z}-\frac{\omega}{v_{\mathrm m}},
\label{eq1}
\end{equation}
where \(k_{z}\) is the normal wave-vector component and \(\omega\) the angular frequency. For a plane wave in medium \(j\) with index \(n_{j}\) and incidence angle \(\phi_{j}\), the angle enters through \(k_{z} = (n_{j}\omega_{j}/ c)\cos\phi_{j}\), and space-time refraction requires equality of \(T\) and \(k_{x}\) on the two sides. Together with the dispersion relations, these constraints determine how the spatio-temporal spectrum of the incident STWP maps to the transmitted one. In the stationary limit \(v_{\mathrm m}\to 0\) they reduce to conservation of transverse momentum and frequency, while in the opposite limit \(v_{\mathrm m}\to \infty\) conservation of \(T\) tends to conservation of \(k_{z}\), corresponding to a purely temporal index jump.

The STWPs that we consider are propagation-invariant pulsed beams obtained by intersecting the light cone with tilted spectral planes, as sketched in Fig. \ref{intro}(d). The tilt angle \(\theta\) of the plane in the \((k_{z},\omega/ c)\) projection sets the group velocity through the group index

\begin{equation}
\tilde n=\frac{ c}{v_{\mathrm g}}=\cot\theta,
\label{eq2}
\end{equation}

\noindent so different spectral planes correspond to different group velocities.

Baseband STWPs resemble conventional pulses built around a carrier frequency inside the band. Spectrally, they arise from a plane that intersects the light cone near the axis, so that the intersection passes through a carrier point with \(k_{x}\approx 0\) and \(\omega\approx\omega_{0}\) [Fig. \ref{intro}(d), left] \cite{yessenov2021refraction, longhi2004gaussian}. In spatial refraction at a time-invariant interface, the associated invariant is \(n(n-\tilde n)\), which encodes the curvature of the spectral locus and determines how the group index transforms \cite{yessenov2021refraction, bhaduri2020anomalous}. Baseband packets can realise subluminal, luminal, or superluminal group velocities while remaining close to an on-axis carrier and provide the closest analogue of familiar pulsed beams.

X--wave STWPs form a second family [Fig. \ref{intro}(d), centre]. They arise from a spectral plane that passes through the origin, so the intersection with the light cone consists of straight lines through the origin in the \((k_{x},\omega/ c)\) projection \cite{yessenov2021refraction, saari1997evidence}. The two branches at \(\pm k_{x}\) combine to form a propagation-invariant X-shaped spatio-temporal profile. In the static case, their refraction is governed by the invariant \(n^{2}-\tilde n^{2}\), which controls how the group index transforms across the interface \cite{yessenov2021refraction}. X--waves are more challenging to synthesise compared to baseband STWPs \cite{yessenov2021refraction}, but exemplify propagation-invariant packets whose spectra are centred away from \(k_{x}=0\) while still obeying a simple linear constraint.

Sideband STWPs constitute a third family [Fig. \ref{intro}(d), right]. Here the spectral plane does not pass through the origin and does not cross the light cone at \(k_{x}=0\); the support lies on a shifted conic segment at finite \(k_{x}\) and finite frequency offset \cite{yessenov2021refraction, brittingham1983focus}. It is convenient to label these packets by a sideband parameter \(\zeta\), which measures the offset along the \(\omega\) axis between the carrier point and the origin, together with an effective group index \(\tilde n\) \cite{yessenov2021refraction}. In spatial refraction, the relevant invariant involves the product \((n+\tilde n)(n-\zeta\tilde n)\), which captures both the slope of the spectral line and the sideband displacement \cite{yessenov2021refraction}. Sideband packets bridge the gap between baseband and X--wave configurations and access parameter regimes that are not available to the other families.

For time-invariant planar interfaces, refraction laws at normal incidence have been derived for all three STWP families, and oblique incidence has been analysed in detail only for baseband packets \cite{yessenov2021refraction}. In that setting the conserved transverse momentum and frequency lead to the oblique invariant \(n(n-\tilde n)\cos^{2}\phi\). This approach does not extend meaningfully to X--wave or sideband packets, whose spectra are not centred at \(k_{x}=0\); their two lobes impinge at different angles and refract asymmetrically, destroying strict propagation invariance except near the luminal limit. A moving interface reinforces this distinction, since conservation of \(k_{x}\) and of the Doppler invariant \(T\) introduces additional skewing for off-axis spectra. We therefore allow arbitrary incidence angles only for baseband STWPs, whose spectra remain tightly localised around \(k_{x}=0\) and admit a controlled small-angle expansion. X--wave and sideband STWPs are treated at normal incidence, where their symmetric spectra remain propagation-invariant. With this framework in place, we now formulate the corresponding space-time refraction laws and examine how they interpolate between spatial and purely temporal limits. A quantitative analysis of the resulting oblique incidence distortion and strategies to suppress it are given in Supplementary Section S1.

\subsection{Baseband space-time refraction law}

For baseband STWPs the spectrum is concentrated near \(k_{x}=0\), so a narrowband paraxial description around the on-axis carrier is appropriate \cite{yessenov2021refraction, yessenov2022space}. As before, we denote by \(n_{1}\) and \(n_{2}\) the phase indices in the two media, by \(\tilde n_{1}\) and \(\tilde n_{2}\) the corresponding group indices of the incident and transmitted packets, and by \(\phi_{1}\) and \(\phi_{2}\) the angles between the packet axes and the interface normal. It is convenient to introduce the dimensionless interface parameter \(\xi =  c/v_{\mathrm m}\), which compares the interface velocity with the vacuum speed of light.

In the narrowband and paraxial regime, conservation of the tangential momentum \(k_{x^{\prime}}\) and of the Doppler invariant \(T(\phi)=k_{z^{\prime}}-\omega/v_{\mathrm m}\) in a frame \((x^{\prime},z^{\prime})\) aligned with the packet axis, together with the dispersion relations on each side, yields an oblique space-time refraction law that couples the incidence angles and the local spectral curvature. The resulting relation

\begin{equation}
\frac{(\xi - \tilde n_{1}\cos\phi_{1})(n_{1}\cos\phi_{1}-\xi)}{n_{1}(\tilde n_{1}-n_{1})\cos^{2}\phi_{1}}
=
\frac{(\xi - \tilde n_{2}\cos\phi_{2})(n_{2}\cos\phi_{2}-\xi)}{n_{2}(\tilde n_{2}-n_{2})\cos^{2}\phi_{2}},
\label{eq3}
\end{equation}

\noindent provides an implicit map from \((\tilde n_{1},\phi_{1})\) to \((\tilde n_{2},\phi_{2})\). At normal incidence, \(\phi_{1}=\phi_{2}=0\), the angular factors drop out and Eq.  \ref{eq3} reduces to

\begin{equation}
\frac{(\xi - \tilde n_{1})(n_{1}-\xi)}{n_{1}(\tilde n_{1}-n_{1})}
=
\frac{(\xi - \tilde n_{2})(n_{2}-\xi)}{n_{2}(\tilde n_{2}-n_{2})},
\label{eq4}
\end{equation}

\noindent which directly relates the incident and transmitted group indices (in agreement with \cite{wang2025spatiotemporal}). The derivation of Eq. \ref{eq3}, including the paraxial expansion, the carrier constraints, and the explicit purely spatial and purely temporal interface limits, is summarised in Supplementary Section S1.

Despite its compact form, Eq.  \ref{eq3} encodes several features that are central for the applications discussed later. For fixed \(n_{1}\) and \(n_{2}\) there is a co-moving value of the group index on each side, \(\tilde n_{j}=\xi/\cos\phi_{j}\), at which the numerator of Eq.  \ref{eq3} vanishes and the mapping leaves the group index essentially unchanged. Away from these co-moving values, the normal incidence law in Eq.  \ref{eq4} can compress a broad range of incident group indices into a narrow range of transmitted values for suitable choices of \(\xi\) between \(n_{1}\) and \(n_{2}\). This parameter regime underlies the ``space-time optical push broom'' effect that we analyse in the applications section, and the associated slope, accumulation behaviour, and co-moving fixed points of the mapping are discussed in detail in Supplementary Section S2.

\subsection{X--wave space-time refraction law}

X--wave STWPs are built from spectral planes through the origin, so their support on the light cone consists of straight lines symmetric in \(k_{x}\). In this case the relations between \((k_{x},k_{z},\omega)\) and the group index \(\tilde n\) are exact and do not rely on a parabolic approximation \cite{yessenov2021refraction, yessenov2022space}. For a moving interface that is uniform along \(x\) and travels along \(z\) with velocity \(v_{\mathrm m}\), conservation of the tangential momentum and of the Doppler invariant \(k_{z}-\omega/v_{\mathrm m}\) leads, together with the dispersion relations, to the X--wave space-time refraction law

\begin{equation}
\frac{\tilde n_{1}-\xi}{\sqrt{n_{1}^{2}-\tilde n_{1}^{2}}}
=
\frac{\tilde n_{2}-\xi}{\sqrt{n_{2}^{2}-\tilde n_{2}^{2}}},
\label{eq5}
\end{equation}

\noindent with existence domain \(n_{j}^{2}-\tilde n_{j}^{2}>0\) and interface parameter \(\xi= c/v_{\mathrm m}\) (see Supplementary Section S1 for the exact derivation and limiting cases). The combination \(n^{2}-\tilde n^{2}\) familiar from static refraction \cite{yessenov2021refraction} now appears under a square root, while \(\xi\) enters additively in the numerator. As in the baseband case, there is a co-moving condition \(\tilde n_{j}=\xi\) on each side where the numerator vanishes and the interface leaves the group index essentially unchanged.

A notable feature of Eq.  \ref{eq5} is that it is quadratic in \(\tilde n_{2}\). For suitable combinations of \((n_{1},n_{2},\tilde n_{1},\xi)\) it admits two distinct physical roots inside the allowed domain, so a single incident X--wave can split into two transmitted X--waves that share the same carrier and spatial profile but propagate with different group velocities. This ``velocity spectral fission'' is specific to the moving interface and has no counterpart in the time-invariant case, and its implications are examined in the applications section. The explicit expressions for the two transmitted branches and the parameter domain for double refraction are given in Supplementary Section S3.

\subsection{Sideband space-time refraction law}

Sideband STWPs occupy spectral regions shifted away from both the origin and from \(k_{x}=0\). They are described by a spectral plane that intersects the light cone on a conic segment at finite \(k_{x}\) and by a sideband parameter \(\zeta\) that quantifies the frequency offset between the carrier and the band edge. In the static problem their refraction at normal incidence is governed by a slope invariant that involves the combination \((n+\tilde n)(n-\zeta\tilde n)\). For a moving interface this slope constraint must be combined with the Doppler invariant, which leads to a modified mapping between incident and transmitted packets.

In the narrowband regime, the sideband space-time law at normal incidence can be written as

\begin{equation}
\frac{(1-\zeta_{2})(n_{1}+\tilde n_{1})(n_{1}-\zeta_{1}\tilde n_{1})}{\tilde n_{2}-\xi}
=
\frac{(1-\zeta_{1})(n_{2}+\tilde n_{2})(n_{2}-\zeta_{2}\tilde n_{2})}{\tilde n_{1}-\xi},
\label{eq7}
\end{equation}

\noindent where \(\zeta_{1}\) and \(\zeta_{2}\) are the sideband parameters of the incident and transmitted packets, \(0<\zeta_{j}<1\), and \(\xi= c/v_{\mathrm m}\). In addition to Eq.  \ref{eq7}, the carrier frequencies obey a separate constraint that links \(\zeta_{1}\) and \(\zeta_{2}\) for given \(n_{j}\) and \(\tilde n_{j}\), so the pair \((\tilde n_{2},\zeta_{2})\) is determined implicitly by the incident parameters and by the interface velocity. The detailed derivation of Eq. \ref{eq7}, including the carrier constraint that links \(\zeta_{1}\) and \(\zeta_{2}\) and its spatial, temporal, and X--wave limits, is provided in Supplementary Section S1.

As \(\zeta_{j}\to 1\), the carrier approaches the band edge and Eq.  \ref{eq7} continuously approaches the X--wave law in Eq.  \ref{eq5}, so sideband packets interpolate smoothly between baseband and X--wave behaviour. Mathematically, Eq.  \ref{eq7} is quadratic in \(\tilde n_{2}\); for suitable \((n_{1},n_{2},\tilde n_{1},\zeta_{1},\xi)\) it can admit two physical roots, corresponding to double refraction into two transmitted sideband packets. The parameter domain for this regime is summarised in Supplementary Section S1. In practice, the narrowband requirement confines \(\zeta\) to values close to unity, so sideband packets remain near luminal group speeds and mainly serve as a continuous bridge between baseband and X--wave regimes.

\section{Applications}

\begin{figure}
 \centering
 \begin{adjustbox}{width=0.8\linewidth, center}
  \includegraphics[keepaspectratio]{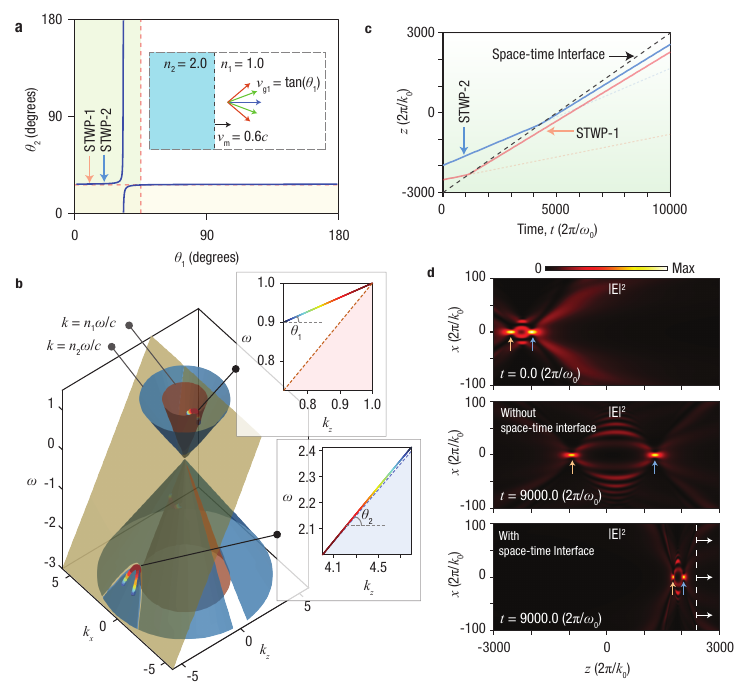}
  \end{adjustbox}
\caption{\textbf{Space-time anomalous optical push broom via refraction of baseband STWPs.}
\textbf{a}, Space-time refraction law for baseband STWPs at normal incidence for \(n_{1}=1\), \(n_{2}=2\), and \(v_{\mathrm m}=0.6 c\). The mapping \(\theta_{1}\!\to\!\theta_{2}\) compresses a wide span of incident group indices into a narrow transmitted range. Orange and blue arrows mark two representative inputs. Inset: geometry of a baseband STWP incident on a planar space-time interface.
\textbf{b}, Spectral construction in \((k_{x},k_{z},\omega)\) space. Red and blue cones are the light cones of media \(1\) and \(2\); the yellow plane is the space-time refraction plane that maps the incident spectral locus to the transmitted one. Insets: \((k_{z},\omega)\) projections showing the change in spectral tilt.
\textbf{c}, World lines of the interface and of the packet centres at \(x=0\). Dashed black: interface trajectory. Solid orange and blue: group trajectories with the interface present. Dotted lines: reference propagation without the interface. The moving boundary sweeps both packets onto nearly the same group velocity.
\textbf{d}, Field intensity snapshots \(\lvert E\rvert^{2}\) in the \((x,z)\) plane. Top: input at \(t=0\). Middle: separation of the two STWPs in a uniform medium. Bottom: co-propagation after crossing the space-time interface. The dashed vertical line marks the interface location, and arrows track the packet peaks (see Supplementary Video 1 for a full time animation).
}
    \label{push_broom}
\end{figure}

\subsection{Space-time optical push broom}

The baseband space-time refraction law at normal incidence, Eq. \ref{eq4}, with \(\xi =  c/v_{\mathrm m}\) and \(\tilde n_{j}= c/v_{\mathrm g j}\), defines a mapping from incident to transmitted group indices that we use to realise a ``space-time optical push broom''. Classical optical push broom schemes arise when a refractive index front moves through a dispersive structure and continuously gathers energy from a weak signal into a narrow region near the front. In fibre Bragg gratings and related guided platforms this leads to strong temporal compression, frequency translation, and large effective delays, which have been proposed and used for optical buffering, pulse shaping, and all-optical switching \cite{de1992optical,broderick1997optical,gaafar2019front,pendry2024air, zhang2025pulse}. In these settings, however, the front velocity is tied to the pump group velocity and the signal group velocity is fixed by material dispersion, so the relative speed between the front and the probe is only weakly tunable. In our space-time platform, baseband STWPs in a homogeneous medium are labelled by a spectral tilt angle, so their group indices \(\tilde n_{j}\) can be tuned almost arbitrarily at fixed carrier frequency, and recent demonstrations of synthetic motion allow the interface velocity \(v_{\mathrm m}\) to be prescribed independently using ultrafast pumps incident at controlled angles \cite{harwood2025space}. This independent control of signal group velocity and interface velocity, combined with the mapping in Eq. \ref{eq4}, enables a space-time optical push broom in which a broad range of incident group velocities is compressed into a narrow band of transmitted velocities in a diffraction-free, waveguide-free setting. A quantitative analysis of the mapping slope, accumulation of incident group indices, and design conditions for efficient push broom operation is presented in Supplementary Section S2.

Figure \ref{push_broom} illustrates this effect for baseband STWPs incident normally on a moving interface between media with \(n_{1}=1\) and \(n_{2}=2\), and interface velocity \(v_{\mathrm m}=0.6  c\) so that \(\xi \simeq 1.67\). For these parameters Eq. \ref{eq4} produces a refraction curve \(\theta_{2}(\theta_{1})\) that is nearly flat over a wide range of incident spectral tilt angles (see Fig. \ref{push_broom}(a)). Many distinct incident group indices in medium-1 are therefore mapped to almost the same transmitted group index in medium-2, which is the defining feature of the push broom regime. Two representative packets, labelled STWP-1 and STWP-2 in Fig. \ref{push_broom}(a), lie in this flat region and have clearly different incident group velocities.

The spectral construction in Fig. \ref{push_broom}(b) makes this mapping geometrical. The light cones of the two media and the space-time refraction plane set by the interface determine a unique transmitted spectral sheet for each incident sheet. For the chosen parameters the frequency ratio

\begin{equation}
\frac{\omega_{2}}{\omega_{1}}
=
\frac{n_{1} - \xi}{n_{2} - \xi},
\label{eq9}
\end{equation}

\noindent is negative, so the transmitted support lies on the negative frequency branch of the \(n_{2}\) cone. Using the equivalence of plane waves at \((k_{x},k_{z},\omega)\) and \((-k_{x},-k_{z},-\omega)\), this can be viewed as refraction to a phase conjugated transverse component, combined with a compressed range of group velocities. The algebra leading to Eq. \ref{eq9} and the associated interpretation in terms of negative frequency and phase conjugation are discussed in Supplementary Section S2.

The real-space trajectories in Fig. \ref{push_broom}(c) show the same effect in the space-time domain. Without the interface, the centres of STWP-1 and STWP-2 separate steadily according to their different group velocities. In the presence of the moving interface, both packets are swept onto world lines that remain close to one another and to the interface trajectory, consistent with the nearly constant transmitted group index predicted by Eq. \ref{eq4}. Notably, this compression occurs in an \textit{anomalous regime} where the transmitted group velocity increases even though the refractive index of medium-2 is higher than that of medium-1, opposite to the behaviour expected for conventional wave packets in time-invariant media. Field snapshots in Fig. \ref{push_broom}(d) confirm that the two packets, which would have separated significantly in a spatially homogeneous time-invariant medium, remain close in space and time after crossing the interface and propagate with nearly matched velocities in medium-2 (for a full time animation see Supplementary Video 1).

These results show that a single moving space-time interface can realise a ``space-time optical push broom'' for baseband STWPs. By tuning the interface velocity and refractive indices according to the refraction law, one can compress a large set of input group velocities into a narrow transmitted band without operating arbitrarily close to luminal conditions, and at the same time access negative-frequency and phase-conjugated branches of the dispersion relation. In contrast to fibre-based implementations, this realisation operates in a homogeneous medium using non-diffractive structured waves and a synthetically driven interface, which is attractive for compact time gating, programmable buffering, and signal routing in space-time photonic platforms.

\begin{figure}
 \centering
 \begin{adjustbox}{width=0.8\linewidth, center}
  \includegraphics[keepaspectratio]{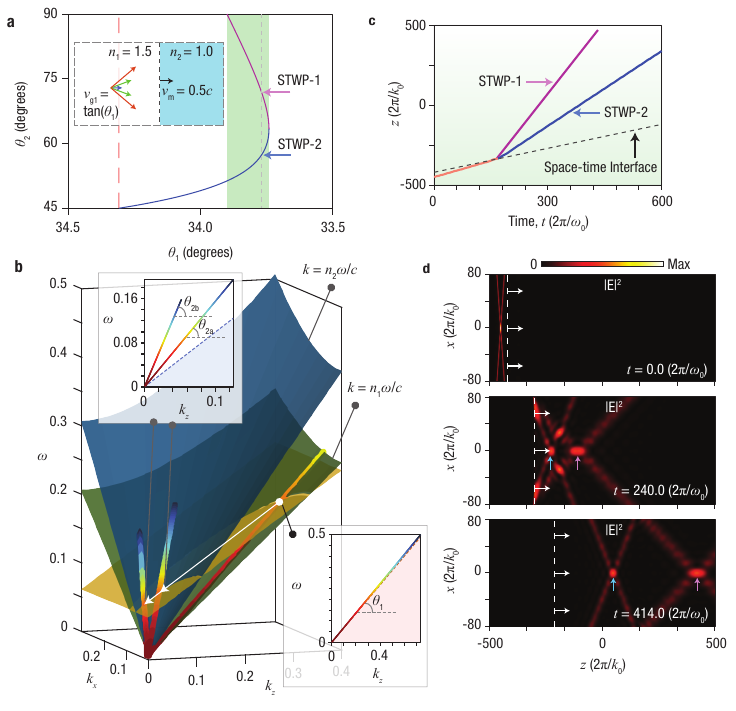}
  \end{adjustbox}
\caption{\textbf{Velocity spectral optical fission of an X--wave at a moving space-time interface.}
\textbf{a}, Space-time refraction law for an X--wave at normal incidence for \(n_{1}=1.5\), \(n_{2}=1.0\), and \(v_{\mathrm m}=0.5 c\). The mapping \(\theta_{1}\!\to\!\theta_{2}\) becomes double valued over a narrow interval of incident tilts (green band), so one input X--wave produces two transmitted branches. Magenta and blue arrows mark two representative outputs. Inset: geometry of an X--wave incident on a moving space-time interface.
\textbf{b}, Spectral construction in \((k_{x},k_{z},\omega)\) space. Green and blue surfaces are the light cones of media \(1\) and \(2\). The yellow plane is the space-time refraction plane, which intersects the \(n_{2}\) cone along two distinct loci corresponding to the two transmitted X--waves. Insets: \((k_{z},\omega)\) projections showing the incident tilt \(\theta_{1}\) and the two transmitted tilts \(\theta_{2a}\) and \(\theta_{2b}\). Only the physical octant \(k_{x},k_{z},\omega>0\) is shown.
\textbf{c}, World lines at \(x=0\). Dashed black: interface trajectory. Orange: incident X--wave. Magenta and blue: centres of the two transmitted packets, which separate according to their distinct group velocities.
\textbf{d}, Field intensity snapshots \(\lvert E\rvert^{2}\) at \(t=0\), \(240\,(2\pi/\omega_{0})\), and \(414\,(2\pi/\omega_{0})\). The vertical dashed line marks the interface. Magenta and blue arrows track STWP-1 and STWP-2, whose spatial separation reflects velocity spectral optical fission (see Supplementary Video 2 for a full time animation).
}
    \label{double_refraction}
\end{figure}

\subsection{Velocity spectral optical fission}

The X--wave space-time refraction law at normal incidence, Eq.  \ref{eq5}, is quadratic in the transmitted group index \(\tilde n_{2}\). For suitable combinations of \((n_{1},n_{2},v_{\mathrm m})\) and incident tilt \(\tilde n_{1}\), this quadratic admits two physical solutions inside the X--wave existence domain. In that regime a single incident X--wave splits into two propagation-invariant transmitted X--waves with distinct group indices and hence distinct group velocities, a process that we refer to as ``velocity spectral optical fission''. The detailed algebra, discriminant analysis, and parameter conditions for double refraction are summarised in Supplementary Section S3.

Figure \ref{double_refraction} illustrates this effect for an interface between media with \(n_{1}=1.5\), \(n_{2}=1.0\), and velocity \(v_{\mathrm m}=0.5  c\). The refraction curve \(\theta_{2}(\theta_{1})\) in Fig. \ref{double_refraction}(a) is single valued over most of the incident tilt range, but develops a narrow interval where two transmitted tilts are allowed and both satisfy the X--wave existence condition. In this interval, highlighted by the green band, one incident X--wave produces two transmitted X--waves, labelled STWP-1 and STWP-2. The corresponding branches have different spectral tilt angles and therefore different group velocities in medium-2.

The geometric origin of the splitting is shown in Fig. \ref{double_refraction}(b). The space-time refraction plane defined by the moving interface intersects the light cone of medium-2 along two distinct curves that both satisfy the X--wave constraint in \((k_{x},k_{z},\omega)\) space. Their projections onto the \((k_{z},\omega)\) plane are straight lines with different tilt angles, confirming that each transmitted branch remains an X--wave with its own constant group velocity. The world lines in Fig. \ref{double_refraction}(c) show how the centre of the incident packet divides into two transmitted trajectories with different slopes: one packet outruns the interface, while the other lags behind, so that the energy is deterministically split into two velocity channels. Field snapshots in Fig. \ref{double_refraction}(d) display two well separated X-shaped intensity profiles after the interface, in agreement with the predicted group velocities (for a full time animation see Supplementary Video 2).

Together, these results demonstrate that a moving space-time interface can realise ``velocity spectral optical fission'' of X--waves, creating two propagation-invariant channels with distinct velocities from a single incident packet. This mechanism suggests compact implementations of velocity multiplexers, programmable delay elements, and multichannel signal processing based on space-time structured media.

\section{Conclusion}\label{sec3}

Space-time refraction of STWPs provides a unified picture for how structured pulsed beams interact with moving interfaces. By identifying the conserved tangential momentum and a Doppler type invariant, we derived refraction laws for baseband, X--wave, and sideband packets that are valid for arbitrary interface velocities and, in the baseband case, for oblique incidence. These laws connect smoothly to the static and purely temporal limits, clarify the role of optical horizons, and place earlier observations of front induced transitions within a single geometric framework.

Within this framework, baseband packets emerge as natural candidates for space-time optical push broom, since their spectra remain narrowly concentrated around the axis even at finite angle. We showed that a moving interface can compress a wide range of incident group velocities into a narrow transmitted band without driving the packet into a singular luminal condition, and that the same law identifies co-moving packets whose group velocity is essentially preserved. X--waves, in turn, support velocity spectral optical fission, in which a single incident packet splits into two propagation-invariant branches with distinct group velocities, while sideband packets form a continuous bridge between baseband and X--wave behaviour. Together, these examples show how a single moving interface can reshape the velocity content of structured fields in ways that are difficult to realise with static components alone.

The present results suggest several directions for further work. On the theoretical side, our treatment assumed propagation-invariant STWPs in the form of light-sheets rather than cylindrically symmetric profiles, and it focused on nondispersive media with an impedance-matched interface. Extending the refraction laws to dispersive or weakly inhomogeneous platforms, and incorporating the angular dependence of the Fresnel coefficients, which can distort finite-aperture packets at large spatial bandwidths or oblique incidence, would bring the description closer to realistic experiments. On the experimental side, baseband STWPs are already well established \cite{yessenov2022space}. X--wave and sideband STWPs, however, remain more challenging to synthesize \cite{yessenov2021refraction}, leaving clear room for experimental advances. Finally, rapid progress in optically driven index fronts \cite{gaafar2019front} and space-time metamaterials \cite{galiffi2022photonics}, including emerging materials platforms such as ITO \cite{harwood2025space}, indicates that controllable space-time interfaces can be implemented in compact devices. In such settings, the refraction mechanisms explored here could support reconfigurable time gating, selective buffering, and velocity multiplexing of STWPs, and provide new routes to probe horizon-like conditions and energy flow in dynamic photonic structures.

\section*{{Methods}}

\bmhead{Numerical simulations} Numerical results were obtained by constructing the scalar field as a superposition of plane waves whose spatio-temporal spectra follow the analytic loci of the chosen STWPs in \((k_{x},k_{z},\omega)\) space. For each configuration the spectrum was discretised on a finite grid, the moving interface between homogeneous nondispersive media with indices \(n_{1}\) and \(n_{2}\) and velocity \(v_{\mathrm m}\) along \(z\) was imposed by enforcing conservation of \(k_{x}\) and of the Doppler invariant \(T = k_{z} - \omega / v_{\mathrm m}\), and the transmitted spectral locus was obtained accordingly. The time-dependent field \(E(x,z,t)\) was then evaluated on a Cartesian \((x,z)\) grid and a set of time samples by summing \(\exp[\mathrm i(k_{x}x + k_{z}z - \omega t)]\) over all spectral components, with impedance matching assumed so that only incident and transmitted waves are included. Baseband, X--wave, and sideband cases differ only in the spectral constraint and parameter choices (carrier frequency, bandwidth, spectral tilt, and angular apodisation), while the numerical scheme is otherwise identical, and snapshots, world lines, and animations are extracted from the resulting field data.

\backmatter

\bmhead{Supplementary information}
Supplementary Sections S1–3.


\section*{Declarations}

\bmhead{Funding}
Z. H. and J. B. P. acknowledge support from the Engineering and Physical Sciences Research Council (EPSRC) under grant EP/Y015673/1.

\bmhead{Conflict of interest/Competing interests}
The authors declare no competing interests.

\bmhead{Availability of data and materials}
Authors confirm that all relevant data are included in the paper and/or its Supplementary Information files.

\bmhead{Code availability}
The code used to produce these results is available upon reasonable request to the corresponding author.

\bibliography{sn-bibliography}

\clearpage
\setcounter{section}{0}
\setcounter{subsection}{0}
\setcounter{subsubsection}{0}
\setcounter{equation}{0}
\setcounter{figure}{0}
\setcounter{table}{0}

\renewcommand{\theequation}{S\arabic{equation}}
\renewcommand{\thesection}{S\arabic{section}}
\renewcommand{\thefigure}{S\arabic{figure}}

\title{
  \begin{center}
    {\Large  \textbf{Supplementary Material:\\ \vspace{5pt} Space-time refraction\\ \vspace{3pt} of space-time wave packets}}
  \end{center}
}

\maketitle
\section{Space-time refraction: derivations}\label{secS1}
\subsection{Baseband space-time refraction law}

We consider the planar interface geometry shown in Fig. 1(b) of the main article, where \((x,z)\) are attached to the interface with \(x\) tangent and \(z\) normal. For the spectral construction it is convenient to introduce a wave aligned frame \((x',z')\) in each homogeneous region, with \(z'\) along the STWP propagation direction and \(x'\) transverse to it. The interface normal \(z\) is seen from the wave aligned axis \(z'\) under an angle \(\phi\). Throughout this subsection \(k_{x'}\) and \(k_{z'}\) denote the transverse and longitudinal components of the wave vector in the STWP aligned frame, while \(k_{x}\) and \(k_{z}\) denote the corresponding components in the interface frame.

A baseband space-time wave packet (STWP) in a medium of refractive index \(n\) is constructed by intersecting the light cone with a spectral plane that crosses the cone at \(\omega_{0}=c\,k_{0}\). In the \((\omega/c,k_{x'})\) and \((k_{z'},k_{x'})\) projections the exact baseband conics are

\begin{equation}
    \Bigl(\frac{\omega}{c}-\frac{\tilde n}{\tilde n+n}k_{0}\Bigr)^{2}
    +\frac{1}{\tilde n^{2}-n^{2}}\,k_{x'}^{2}
    =\frac{n^{2}}{(\tilde n+n)^{2}}k_{0}^{2},
\label{eq:bb-omega-conic}
\end{equation}

\begin{equation}
    \Bigl(\frac{k_{z'}}{\tilde n}-\frac{n^{2}}{\tilde n(\tilde n+n)}k_{0}\Bigr)^{2}
    +\frac{1}{\tilde n^{2}-n^{2}}\,k_{x'}^{2}
    =\frac{n^{2}}{(\tilde n+n)^{2}}k_{0}^{2}.
\label{eq:bb-kz-conic}
\end{equation}

We work in the paraxial regime \(\lvert k_{x'}\rvert\ll k_{0}\) and expand the physical branches that approach \(\omega\to c\,k_{0}\) and \(k_{z'}\to n\,k_{0}\):

\begin{equation}
    \omega \approx c\,k_{0}-\alpha\,k_{x'}^{2}, \qquad
    \alpha=\frac{c}{2k_{0}}\frac{1}{n(\tilde n-n)},
\label{eq:bb-omega-exp}
\end{equation}

\begin{equation}
    k_{z'} \approx n\,k_{0}-\beta\,k_{x'}^{2}, \qquad
    \beta=\frac{\tilde n}{2k_{0}}\frac{1}{n(\tilde n-n)}.
\label{eq:bb-kz-exp}
\end{equation}

For an oblique, translating interface the conserved quantities are most naturally written in the interface frame. If \(\phi\) is the angle between the packet axis \(z'\) and the interface normal \(z\), the components of the wave vector in the two frames are related by

\begin{equation}
    k_{x} = k_{x'}\cos\phi - k_{z'}\sin\phi, \qquad
    k_{z} = k_{x'}\sin\phi + k_{z'}\cos\phi.
\end{equation}

\noindent Uniformity along \(x\) enforces conservation of the exact tangential component,

\begin{equation}
    k_{x}\;\;\text{is conserved},
\end{equation}

\noindent while invariance of the co-moving normal coordinate \(s=z-v_{\mathrm m} t\) enforces conservation of the Doppler invariant

\begin{equation}
    \mathcal T(\phi)\equiv k_{z}-\frac{\omega}{v_{\mathrm m}}
    = k_{x'}\sin\phi + k_{z'}\cos\phi - \frac{\omega}{v_{\mathrm m}}.
\label{eq:T-def}
\end{equation}

\noindent These two invariants, together with Eqs. \ref{eq:bb-omega-exp}–\ref{eq:bb-kz-exp}, generate the baseband space-time refraction law quoted in the main text.

\paragraph{Carrier constraints and reduced tangential invariant.}

At the carrier the baseband packet has \(k_{x',0}=0\), \(\omega_{0}=c\,k_{0}\), and \(k_{z',0}=n\,k_{0}\). Using the rotation between the wave aligned frame \((x',z')\) and the interface frame \((x,z)\),

\begin{equation}
    k_{x} = k_{x'}\cos\phi - k_{z'}\sin\phi, \qquad
    k_{z} = k_{x'}\sin\phi + k_{z'}\cos\phi,
\end{equation}

\noindent the interface frame components at the carrier are

\begin{equation}
    k_{x,0} = -\,n\,k_{0}\sin\phi, \qquad
    k_{z,0} = n\,k_{0}\cos\phi.
\end{equation}

\noindent Conservation of the tangential component \(k_{x}\) across the interface therefore yields the carrier constraint

\begin{equation}
    n_{1}k_{0,1}\sin\phi_{1} = n_{2}k_{0,2}\sin\phi_{2}.
\end{equation}

\noindent Likewise, evaluating the Doppler invariant \(k_{z}-\omega/v_{\mathrm m}\) at the carrier gives

\begin{equation}
    \bigl.k_{z}-\omega/v_{\mathrm m}\bigr|_{k_{x',0}=0}
    = k_{0}\!\left(n\cos\phi-\frac{c}{v_{\mathrm m}}\right),
\end{equation}

\noindent and equality across the interface implies

\begin{equation}
    k_{0,1}\Bigl(n_{1}\cos\phi_{1}-\frac{c}{v_{\mathrm m}}\Bigr)
    =
    k_{0,2}\Bigl(n_{2}\cos\phi_{2}-\frac{c}{v_{\mathrm m}}\Bigr).
    \label{eq:const-match}
\end{equation}

\noindent To expose the small, \(k_{x'}\) dependent content of the tangential invariant, write

\begin{equation}
    k_{x} = k_{x,0} + \Delta k_{x},
\end{equation}

\noindent with

\begin{equation}
    \Delta k_{x} \equiv k_{x'}\cos\phi - \bigl(k_{z'}-n\,k_{0}\bigr)\sin\phi.
\end{equation}

\noindent Using the paraxial baseband expansions \ref{eq:bb-omega-exp}–\ref{eq:bb-kz-exp},

\begin{equation}
    k_{z'} \approx n\,k_{0}-\beta \,k_{x'}^{2},
\end{equation}

\noindent gives

\begin{equation}
    \Delta k_{x}
    =
    k_{x'}\cos\phi
    +\beta\,\sin\phi\,k_{x'}^{2}
    +\mathcal O\!\bigl(k_{x'}^{3}\bigr).
\end{equation}

\noindent Exact tangential invariance implies \(k_{x}^{(1)}=k_{x}^{(2)}\). Subtracting the carrier equality \(k_{x,0}^{(1)}=k_{x,0}^{(2)}\) eliminates the large terms proportional to \(n\,k_{0}\sin\phi\) and yields

\begin{equation}
    k_{x',1}\cos\phi_{1}+\beta_{1}\,\sin\phi_{1}\,k_{x',1}^{2}
    =
    k_{x',2}\cos\phi_{2}+\beta_{2}\,\sin\phi_{2}\,k_{x',2}^{2}
    +\mathcal O\!\bigl(k_{x'}^{3}\bigr).
\end{equation}

\noindent Hence, to leading order,

\begin{equation}
    k_{x',1}\cos\phi_{1} \approx k_{x',2}\cos\phi_{2}.
\end{equation}

\noindent Motivated by this reduction we introduce the projected small variable

\begin{equation}
    \kappa \equiv k_{x'}\cos\phi,
    \label{eq:kappa-def}
\end{equation}

\noindent which equals the leading order deviation of \(k_{x}\) from its carrier value in the paraxial limit.

\paragraph{Paraxial form of the Doppler invariant.}

Combining the baseband expansions \ref{eq:bb-omega-exp}–\ref{eq:bb-kz-exp} in the wave aligned frame \((x',z')\) with the rotation to the interface frame,

\begin{equation}
    k_{x} = k_{x'}\cos\phi - k_{z'}\sin\phi, 
    \qquad
    k_{z} = k_{x'}\sin\phi + k_{z'}\cos\phi,
\end{equation}

\noindent and inserting into \ref{eq:T-def} gives, after straightforward algebra,

\begin{equation}
    \mathcal T(\phi)
    = k_{0}\Bigl(n\cos\phi-\frac{c}{v_{\mathrm m}}\Bigr)
    + k_{x'}\sin\phi
    + \Bigl(\frac{\alpha}{v_{\mathrm m}}-\beta\cos\phi\Bigr)\,k_{x'}^{2}
    + \mathcal O\!\bigl(k_{x'}^{3}\bigr).
\end{equation}

\noindent Using the projected variable \(\kappa\) defined in \ref{eq:kappa-def} and \(k_{x'}=\kappa/\cos\phi+\mathcal O(\kappa^{2})\) yields

\begin{equation}
    \mathcal T(\phi)
    = k_{0}\Bigl(n\cos\phi-\frac{c}{v_{\mathrm m}}\Bigr)
    + \kappa\tan\phi
    + \frac{\alpha/v_{\mathrm m}-\beta\cos\phi}{\cos^{2}\phi}\,\kappa^{2}
    + \mathcal O\!\bigl(\kappa^{3}\bigr).
\end{equation}

\noindent The term linear in \(\kappa\) is odd under \(\kappa\mapsto-\kappa\) and therefore does not affect curvature matching between the two symmetric spectral lobes of a baseband STWP. Retaining the even part, which controls the mapping of the spectral envelope, we write

\begin{equation}
    \mathcal T(\phi)\approx k_{0}\Bigl(n\cos\phi-\frac{c}{v_{\mathrm m}}\Bigr)+\frac{\alpha/v_{\mathrm m}-\beta\cos\phi}{\cos^{2}\phi}\,\kappa^{2}.
    \label{eq:T-parax}
\end{equation}

\noindent Let medium \(j\in\{1,2\}\) have parameters \((n_{j},\tilde n_{j},k_{0,j},\phi_{j})\). Continuity of the constant terms in \ref{eq:T-parax} at \(\kappa=0\) gives Eq. \ref{eq:const-match}. Equality of the \(\kappa^{2}\) curvatures gives

\begin{equation}
    \frac{1}{2k_{0,1}}\frac{\tfrac{c}{v_{\mathrm m}}-\tilde n_{1}\cos\phi_{1}}{n_{1}(\tilde n_{1}-n_{1})\cos^{2}\phi_{1}}
    =
    \frac{1}{2k_{0,2}}\frac{\tfrac{c}{v_{\mathrm m}}-\tilde n_{2}\cos\phi_{2}}{n_{2}(\tilde n_{2}-n_{2})\cos^{2}\phi_{2}}.
    \label{eq:curv-match}
\end{equation}

\noindent Eliminating the ratio \(k_{0,2}/k_{0,1}\) with Eq. \ref{eq:const-match} yields the oblique space-time refraction law for baseband STWPs,

\begin{equation}
    \frac{\bigl(\tfrac{c}{v_{\mathrm m}}-\tilde n_{1}\cos\phi_{1}\bigr)\bigl(n_{1}\cos\phi_{1}-\tfrac{c}{v_{\mathrm m}}\bigr)}{n_{1}(\tilde n_{1}-n_{1})\cos^{2}\phi_{1}}
    =
    \frac{\bigl(\tfrac{c}{v_{\mathrm m}}-\tilde n_{2}\cos\phi_{2}\bigr)\bigl(n_{2}\cos\phi_{2}-\tfrac{c}{v_{\mathrm m}}\bigr)}{n_{2}(\tilde n_{2}-n_{2})\cos^{2}\phi_{2}}.
    \label{eq:bb-ST-oblique}
\end{equation}

\paragraph{Checks and limits.}

Normal incidence, \(\phi_{1}=\phi_{2}=0\), reduces Eq. \ref{eq:bb-ST-oblique} to the usual normal incidence form. In the spatial limit, \(v_{\mathrm m}\to 0\), Eq. \ref{eq:bb-ST-oblique} reduces to

\begin{equation}
    n_{1}(n_{1}-\tilde n_{1})\cos^{2}\phi_{1}=n_{2}(n_{2}-\tilde n_{2})\cos^{2}\phi_{2},
\end{equation}

\noindent the stationary oblique law. In the temporal limit, \(v_{\mathrm m}\to\infty\), Eq. \ref{eq:bb-ST-oblique} becomes

\begin{equation}
    \frac{n_{1}-\tilde n_{1}}{\tilde n_{1}}=\frac{n_{2}-\tilde n_{2}}{\tilde n_{2}}.
\end{equation}

\noindent The paraxial expansion assumes \(|k_{x'}|\ll k_{0}\) and excludes the degeneracy \(\tilde n=\pm n\). The projected variable \(\kappa\) requires \(\cos\phi\neq 0\). The curvature coefficient in Eq. \ref{eq:T-parax} changes sign at \(v_{\mathrm m}=c/(\tilde n\cos\phi)\), which marks a transition in the mapping across the interface.

\paragraph{Oblique incidence distortion at a moving boundary (origin, scaling, and suppression).}

At an oblique moving interface the exact invariants are the tangential component \(k_{x}\) and the Doppler combination

\begin{equation}
    \mathcal T(\phi)\equiv k_{z}-\frac{\omega}{v_{\mathrm m}},
\end{equation}

\noindent with \(x\) tangent and \(z\) normal to the interface as in Fig. 1(b) of the main article. To analyse how these constraints act on a baseband STWP, it is convenient to expand \(\mathcal T(\phi)\) in the wave aligned frame \((x',z')\), where \(z'\) follows the packet axis and \(x'\) is transverse. Using the rotation relations between \((x',z')\) and \((x,z)\),

\begin{equation}
    k_{x} = k_{x'}\cos\phi - k_{z'}\sin\phi, \qquad
    k_{z} = k_{x'}\sin\phi + k_{z'}\cos\phi,
\end{equation}

\noindent together with the paraxial baseband expansions

\begin{equation}
    \omega \approx c\,k_{0}-\alpha\,k_{x'}^{2}, 
    \qquad
    k_{z'} \approx n\,k_{0}-\beta\,k_{x'}^{2},
\end{equation}

\noindent one finds in each homogeneous region

\begin{equation}
    \mathcal T(\phi)
    = k_{0}\Bigl(n\cos\phi-\frac{c}{v_{\mathrm m}}\Bigr)
    \;+\; k_{x'}\sin\phi
    \;+\; \Bigl(\tfrac{\alpha}{v_{\mathrm m}}-\beta\cos\phi\Bigr)\,k_{x'}^{2}
    \;+\;\mathcal O(k_{x'}^{3}),
    \label{eq:Tfull-odd-even}
\end{equation}

\noindent which is a small \(k_{x'}\) expansion around the carrier. In terms of the projected variable defined in Eq. \ref{eq:kappa-def},

\begin{equation}
    \kappa \equiv k_{x'}\cos\phi,
\end{equation}

\noindent Eq. \ref{eq:Tfull-odd-even} becomes

\begin{equation}
    \mathcal T(\phi)
    = k_{0}\Bigl(n\cos\phi-\frac{c}{v_{\mathrm m}}\Bigr)
    \;+\; \kappa\tan\phi
    \;+\; \frac{\alpha/v_{\mathrm m}-\beta\cos\phi}{\cos^{2}\phi}\,\kappa^{2}
    \;+\;\mathcal O(\kappa^{3}).
    \label{eq:Tfull-kappa-odd}
\end{equation}

\noindent The odd term \(\kappa\tan\phi\) controls how the two spectral lobes at \(\kappa=\pm\lvert\kappa\rvert\) respond at oblique incidence. For a given pair in Region-1, enforcing both invariants across a jump \(1\!\to\!2\) and using the leading order tangential relation \(k_{x',1}\cos\phi_{1}\simeq k_{x',2}\cos\phi_{2}\Rightarrow \kappa_{1}\simeq\kappa_{2}\) produces equal and opposite first order corrections to the refracted quantities \((\omega_{2},k_{z,2})\) for the \(+\kappa\) and \(-\kappa\) lobes. The resulting separation in the Region-2 longitudinal cut

\begin{equation}
    k_{z',2} = \mathbf k_{2}\!\cdot\!\hat{\mathbf d}_{2},
\end{equation}

\noindent where \(\hat{\mathbf d}_{2}\) is the Region-2 propagation direction set by the carrier invariants, scales as

\begin{equation}
    \bigl|\Delta k_{z',2}\bigr|\;,\ \bigl|\Delta\omega_{2}\bigr|
    \;=\;\mathcal O\!\Bigl(\lvert\kappa\rvert\,\bigl|\tan\phi_{2}-\tan\phi_{1}\bigr|\Bigr),
    \label{eq:odd-scaling}
\end{equation}

\noindent so the splitting is odd in \(\kappa\) and proportional to the difference between \(\tan\phi_{1}\) and \(\tan\phi_{2}\). This behaviour is specific to a moving interface, since it originates from the Doppler term \(-\,\omega/v_{\mathrm m}\) in \(\mathcal T(\phi)\) and vanishes at normal incidence.

\smallskip
\smallskip
\noindent\emph{How to suppress the distortion.} Equation \ref{eq:odd-scaling} makes the remedies clear:

\begin{enumerate}
\item \textbf{Near normal incidence:} choose \(\phi_{1}\) small. The carrier constraints then enforce a small \(\phi_{2}\) as well, and \(|\tan\phi_{2}-\tan\phi_{1}|\ll 1\).
\item \textbf{Temporal regime:} increase \(v_{\mathrm m}\) (temporal jump limit). As \(v_{\mathrm m}\!\to\!\infty\), the carrier angles equalise and the odd term vanishes, \(\tan\phi_{2}\!\to\!\tan\phi_{1}\), so \(|\tan\phi_{2}-\tan\phi_{1}|\!\to\!0\).
\item \textbf{Narrow transverse support:} reduce \(|\kappa|\) (narrower \(k_{x'}\) band). The odd splitting is linear in \(|\kappa|\).
\item \textbf{Stay away from co-moving resonances:} avoid \(v_{\mathrm m}=c/(\tilde n_{j}\cos\phi_{j})\). Near these values the even (quadratic) curvature term flips sign, so any residual imbalance between the two sides can be amplified.
\end{enumerate}

\noindent The even in \(\kappa\) broadening from the quadratic term in Eq. \ref{eq:Tfull-kappa-odd} is still present; it is minimised when the curvature match in Eq. \ref{eq:curv-match} holds, which smooths the envelope mapping across the boundary. The odd skew discussed here is a distinct, purely oblique and space-time effect linked to Doppler mixing.

\paragraph{Comparison with a stationary (purely spatial) boundary.}

For a stationary interface (\(v_{\mathrm m}=0\)) the exact invariants are \(k_{x}\) and \(\omega\). There is no co-moving term and hence no Doppler coupling. With \(\omega\) fixed, the longitudinal component on the refracted side is determined by the light cone constraint

\begin{equation}
    k_{z,2}=\sqrt{\Bigl(\tfrac{n_{2}\omega}{c}\Bigr)^{2}-k_{x}^{2}},
\end{equation}

\noindent so its dependence on \(k_{x}\) is even: expanding about the carrier yields

\begin{equation}
    k_{z,2}=k_{z,2}^{(0)}-\frac{k_{x}^{2}}{2k_{z,2}^{(0)}}+\cdots
\end{equation}

\noindent with no linear term. Consequently, the two transverse lobes \(\pm\kappa\) remain symmetric to first order at oblique incidence; any residual asymmetry appears only at order \(\kappa^{2}\) through curvature differences. In contrast, a moving boundary injects the linear, odd in \(\kappa\) term in Eq. \ref{eq:Tfull-kappa-odd}, producing an \(\mathcal O(|\kappa|)\) lobe separation as in Eq. \ref{eq:odd-scaling}. For the same \((n_{1},n_{2},\phi_{1})\), oblique incidence distortion is therefore generically stronger for a space-time (moving) boundary than for a purely spatial (stationary) boundary, except in the special limits where the odd term vanishes: normal incidence \((\phi_{1}=\phi_{2}=0)\) and the temporal jump regime \((v_{\mathrm m}\to\infty)\).

\subsection{X--wave space-time refraction law}

An X--wave STWP is obtained by intersecting the light cone with a spectral plane that passes through the origin (see Fig. 1(d) middle panel in the main article). In a homogeneous medium of refractive index \(n\), the light cone is \(k_x^2+k_z^2=(n\,\omega/c)^2\). The through--origin spectral plane can be written as \(\omega/c=k_z\tan\theta\), or equivalently \(k_z=\tilde n\,\omega/c\) with \(\tilde n=\cot\theta\). Substituting into the cone gives \(k_x^2=(n^2-\tilde n^2)(\omega/c)^2\). Along the X--wave spectral support the exact relations are therefore

\begin{equation}
    k_z=\tilde n\,\frac{\omega}{c}, \qquad
    k_x=\pm\sqrt{n^2-\tilde n^2}\,\frac{\omega}{c},
    \label{eq:x-support}
\end{equation}

\noindent with the existence condition \(n^2-\tilde n^2>0\).

Consider now a planar interface that is uniform in \(x\) and moves along \(z\) with speed \(v_{\mathrm m}\). Matching phases of fields proportional to \(\exp[i(k_x x+k_z z-\omega t)]\) on the boundary \(z-v_{\mathrm m} t=\text{const}\) yields two linear invariants:

\begin{equation}
    k_{x1}=k_{x2}, \qquad
    k_{z1}-\frac{\omega_1}{v_{\mathrm m}}=k_{z2}-\frac{\omega_2}{v_{\mathrm m}}.
    \label{eq:x-invariants}
\end{equation}

\noindent Inserting Eq. \ref{eq:x-support} gives

\begin{equation}
    \sqrt{n_1^2-\tilde n_1^2}\,\frac{\omega_1}{c}
    =
    \sqrt{n_2^2-\tilde n_2^2}\,\frac{\omega_2}{c},
    \qquad
    \Bigl(\tilde n_1-\frac{c}{v_{\mathrm m}}\Bigr)\frac{\omega_1}{c}
    =
    \Bigl(\tilde n_2-\frac{c}{v_{\mathrm m}}\Bigr)\frac{\omega_2}{c}.
    \label{eq:x-ratios}
\end{equation}

Eliminating \((\omega_1/c)/(\omega_2/c)\) yields the space-time refraction law for X--waves:

\begin{equation}
    \frac{\tilde n_1-\dfrac{c}{v_{\mathrm m}}}{\sqrt{\,n_1^{2}-\tilde n_1^{2}\,}}
    =
    \frac{\tilde n_2-\dfrac{c}{v_{\mathrm m}}}{\sqrt{\,n_2^{2}-\tilde n_2^{2}\,}}.
    \label{eq:x-st-law}
\end{equation}

\noindent This relation is \textit{exact} for X--waves because the support lines are linear in \((\omega/c,k_x,k_z)\) and pass through the origin, so there is no carrier offset and no small--angle expansion.

\paragraph{Multiplicity and quadratic form.}

Define \(t_j\equiv \tilde n_j\) and \(a\equiv c/v_{\mathrm m}\). From Eq. \ref{eq:x-st-law} one has \((t_2-a)=K\sqrt{n_2^2-t_2^2}\) with
\[
K=\frac{t_1-a}{\sqrt{n_1^2-t_1^2}}.
\]

\noindent Squaring and rearranging give a quadratic for \(t_2\),

\begin{equation}
    (1+K^2)\,t_2^2-2a\,t_2+\bigl(a^2-K^2 n_2^2\bigr)=0,
    \label{eq:x-quadratic}
\end{equation}

\noindent with roots

\begin{equation}
    t_{2,\pm}=\frac{2a\pm \sqrt{\,4K^2\bigl((1+K^2)\,n_2^2-a^2\bigr)\,}}{2(1+K^2)}.
    \label{eq:x-roots}
\end{equation}

Reality requires \((1+K^2)\,n_2^2-a^2\ge 0\). Physical acceptability further requires \(n_2^2-t_{2,\pm}^2>0\) and the unsquared sign condition \((t_{2,\pm}-a)\,K\ge 0\). On the forward branch \(0<t_{2,\pm}<n_2\) there can be two physical solutions if \(a>n_2\). In that case the mapping \(t_2\mapsto (t_2-a)/\sqrt{n_2^2-t_2^2}\) is nonmonotonic on \(0<t_2<n_2\) with a turning point \(t^\star=n_2^2/a\). Two distinct forward solutions exist when
\[
-\frac{a}{n_2}<K<-\sqrt{\Bigl(\frac{a}{n_2}\Bigr)^2-1},
\]
the two solutions coalesce at \(K=-\sqrt{(a/n_2)^2-1}\) with \(t_{2,+}=t_{2,-}=t^\star\), and at most one solution exists for \(a\le n_2\). The special case \(K=0\) enforces \(t_2=a\), which is admissible only if \(a<n_2\).

\paragraph{Checks and limits.}

Spatial limit \(v_{\mathrm m}\to 0\): Eq. \ref{eq:x-st-law} reduces to

\begin{equation}
    n_1^2-\tilde n_1^2=n_2^2-\tilde n_2^2,
\end{equation}

\noindent the stationary X--wave refraction invariant. Temporal limit \(v_{\mathrm m}\to\infty\): Eq. \ref{eq:x-st-law} becomes

\begin{equation}
    \frac{\tilde n_1}{\sqrt{n_1^2-\tilde n_1^2}}
    =
    \frac{\tilde n_2}{\sqrt{n_2^2-\tilde n_2^2}},
\end{equation}

\noindent the temporal--jump analogue. If \(v_{\mathrm m}=c/\tilde n_{j}\) on one side, the numerator in Eq. \ref{eq:x-st-law} vanishes and it must vanish on the other side as well (co-moving resonance). The denominator requires \(n_j^2-\tilde n_j^2>0\), so the luminal limit \(\tilde n_j\to n_j\) is singular and should be treated with a plane--wave model. Because X--waves do not have a unique carrier ray, an oblique--incidence generalization is not meaningful \cite{yessenov2021refraction}; the normal--incidence statement above is the physically relevant one.

\subsection{Sideband space-time refraction law}

A sideband STWP is formed by intersecting the light cone with a spectral plane that does not pass through the origin. The plane is
\[
\omega=\omega_0+\frac{c}{\tilde n}\,(k_z+n k_0),\qquad
k_0=\frac{\omega_0}{c},\ \tilde n=\cot\theta .
\]
Define the carrier \(\omega_c>\omega_0\) and \(k_c=\omega_c/c\). Let the carrier components satisfy \(k_{cx}^2+k_{cz}^2=n^2 k_c^2\). Introduce the sideband parameter \(\zeta=1-\omega_0/\omega_c\in\bigl(\tfrac{n}{n+\tilde n},1\bigr)\).

\noindent In the narrowband limit, linearizing the exact conic about the carrier yields the slope invariant on the \((k_x,\omega/c)\) plane \cite{yessenov2021refraction}

\begin{equation}
    (n+\tilde n)\,(n-\zeta\,\tilde n)
    =\frac{k_x'}{\Omega'/c}\,\frac{k_{cx}}{k_c},
    \label{eq:sb-slope}
\end{equation}

\noindent with \(\Omega'=\omega-\omega_c\) and \(k_x'=k_x-k_{cx}\). The space-time spectral plane fixes the \((k_z,\omega/c)\) slope,

\begin{equation}
    k_z'=\tilde n\,\frac{\Omega'}{c}.
    \label{eq:sb-kzprime}
\end{equation}

Consider a planar interface that is uniform in \(x\) and moves along \(z\) with speed \(v_{\mathrm m}\). Translational symmetry in \(x\) conserves \(k_x\). Invariance of the co--moving coordinate \(z-v_{\mathrm m} t\) conserves \(k_z-\omega/v_{\mathrm m}\). We enforce these invariants at two levels: first at zero order on the carriers, which yields a carrier constraint, then at first order on the variations \((k_x',\Omega')\), which yields the refraction law. Using \(k_{cz}=-(n+\tilde n)k_0+\tilde n k_c\) and \(k_c=k_0/(1-\zeta)\), the carrier constraint becomes

\begin{equation}
    \frac{\tilde n_1-\dfrac{c}{v_{\mathrm m}}}{1-\zeta_1}
    -
    \frac{\tilde n_2-\dfrac{c}{v_{\mathrm m}}}{1-\zeta_2}
    =
    (n_1+\tilde n_1)-(n_2+\tilde n_2).
    \label{eq:sb-carrier}
\end{equation}

\noindent This relation links \(\zeta_1\) and \(\zeta_2\) for given \(n_j\) and \(\tilde n_j\), so they are not independent. Since \(k_{c,j}=k_0/(1-\zeta_j)\), one has \(1-\zeta_j>0\) throughout the sideband window.

Apply the invariants to the first--order variations. Conservation of \(k_x\) implies \(k_{x,1}'=k_{x,2}'\), which with Eq. \ref{eq:sb-slope} gives

\begin{equation}
    k_{c,1}(n_1+\tilde n_1)(n_1-\zeta_1\tilde n_1)\,\frac{\Omega'_1}{c}
    =
    k_{c,2}(n_2+\tilde n_2)(n_2-\zeta_2\tilde n_2)\,\frac{\Omega'_2}{c}.
    \label{eq:sb-kxmatch}
\end{equation}

\noindent Conservation of \(k_z-\omega/v_{\mathrm m}\) at first order and Eq. \ref{eq:sb-kzprime} give

\begin{equation}
    \Bigl(\tilde n_1-\frac{c}{v_{\mathrm m}}\Bigr)\frac{\Omega'_1}{c}
    =
    \Bigl(\tilde n_2-\frac{c}{v_{\mathrm m}}\Bigr)\frac{\Omega'_2}{c}.
    \label{eq:sb-Kmatch}
\end{equation}

\noindent Divide Eq. \ref{eq:sb-kxmatch} by Eq. \ref{eq:sb-Kmatch} and use \(k_{c,j}=k_0/(1-\zeta_j)\). After cancelling common factors this yields the space-time refraction law for sideband STWPs.

\begin{equation}
    (1-\zeta_2)\,(n_1+\tilde n_1)(n_1-\zeta_1\tilde n_1)\Bigl(\tilde n_2-\frac{c}{v_{\mathrm m}}\Bigr)
    =
    (1-\zeta_1)\,(n_2+\tilde n_2)(n_2-\zeta_2\tilde n_2)\Bigl(\tilde n_1-\frac{c}{v_{\mathrm m}}\Bigr).
    \label{eq:sb-st-law}
\end{equation}

\paragraph{Multiplicity and quadratic form.}
Equation \ref{eq:sb-st-law} is quadratic in \(\tilde n_2\), hence for a moving boundary up to two mathematical solutions can arise. Define
\[
\begin{array}{l}
A_1\equiv (n_1+\tilde n_1)(n_1-\zeta_1\tilde n_1),\qquad
A_2(\tilde n_2)\equiv (n_2+\tilde n_2)(n_2-\zeta_2\tilde n_2),\\[6pt]
B\equiv (1-\zeta_1)(\tilde n_1-a),\qquad
a\equiv \frac{c}{v_{\mathrm m}}.
\end{array}
\]
Rearranging \ref{eq:sb-st-law} into \( (1-\zeta_2)A_1(\tilde n_2-a)=B\,A_2(\tilde n_2) \) gives

\begin{equation}
    B\zeta_2\,\tilde n_2^2+(1-\zeta_2)\bigl(A_1-B n_2\bigr)\tilde n_2-\bigl(B n_2^2+(1-\zeta_2)A_1 a\bigr)=0,
    \label{eq:sb-quadratic}
\end{equation}

\noindent so that two physical roots may exist within the sideband window provided Eq. \ref{eq:sb-carrier} is satisfied. The co-moving resonance \(v_{\mathrm m}=c/\tilde n_j\) makes \(\tilde n_j-\tfrac{c}{v_{\mathrm m}}=0\) and requires special care in analysis and numerics.

\paragraph{Checks and limits.}

Spatial limit \(v_{\mathrm m}\to 0\): both factors \(\tilde n_j-\tfrac{c}{v_{\mathrm m}}\) become equal and large, and Eq. \ref{eq:sb-st-law} reduces to

\begin{equation}
    (1-\zeta_2)\,(n_1+\tilde n_1)(n_1-\zeta_1\tilde n_1)
    =
    (1-\zeta_1)\,(n_2+\tilde n_2)(n_2-\zeta_2\tilde n_2).
\end{equation}

\noindent For a stationary boundary the frequency is conserved, so \(\zeta_1=\zeta_2\), which recovers the spatial sideband refraction law

\begin{equation}
    (n_1+\tilde n_1)(n_1-\zeta\,\tilde n_1)
    =
    (n_2+\tilde n_2)(n_2-\zeta\,\tilde n_2).
\end{equation}

\noindent Temporal limit \(v_{\mathrm m}\to\infty\): Eq. \ref{eq:sb-st-law} becomes

\begin{equation}
    (1-\zeta_2)\,(n_1+\tilde n_1)(n_1-\zeta_1\tilde n_1)\,\tilde n_2
    =
    (1-\zeta_1)\,(n_2+\tilde n_2)(n_2-\zeta_2\tilde n_2)\,\tilde n_1,
\end{equation}

\noindent and Eq. \ref{eq:sb-carrier} reduces to

\begin{equation}
    \frac{\tilde n_1}{1-\zeta_1}-\frac{\tilde n_2}{1-\zeta_2}
    =(n_1+\tilde n_1)-(n_2+\tilde n_2).
\end{equation}

X--wave limit \(\zeta\to 1\): with \(A_j\equiv(n_j+\tilde n_j)(n_j-\zeta_j\tilde n_j)\to n_j^2-\tilde n_j^2\) and \(1-\zeta_j\to 0\), keeping the leading nonvanishing order in Eqs. \ref{eq:sb-st-law} and \ref{eq:sb-carrier} and eliminating the vanishing prefactors recovers the X--wave space-time law. Parameter range: \(\zeta\in\bigl(\tfrac{n}{n+\tilde n},1\bigr)\) with \(\zeta=\dfrac{1+\eta}{1+\tilde n/n}\) and \(0<\eta<1\). The narrowband requirement typically forces \(\zeta\to 1\), so sideband packets remain very close to luminal group speed. Oblique incidence is not formulated for sideband packets because the off--axis spectrum produces two impinging lobes that mix tangential and normal components; the normal--incidence law above is the relevant statement.

\section{Space-time optical push broom}

\subsection{Baseband space-time refraction law as a push broom mapping}

For baseband STWPs incident normally on a moving interface, conservation of the tangential momentum and of the Doppler invariant \(k_{z}-\omega/v_{\mathrm m}\) leads to the baseband refraction law

\begin{equation}
    \frac{(\xi - \tilde n_{1})(n_{1} - \xi)}{n_{1}(\tilde n_{1} - n_{1})}
    =
    \frac{(\xi - \tilde n_{2})(n_{2} - \xi)}{n_{2}(\tilde n_{2} - n_{2})},
    \label{eq:S_baseband-normal}
\end{equation}

\noindent where \(n_{1}\) and \(n_{2}\) are the phase indices, \(\tilde n_{j}=c/v_{\mathrm g j}\) are the group indices of the incident and transmitted STWPs, and \(\xi=c/v_{\mathrm m}\) encodes the interface velocity. In the narrowband approximation this mapping applies to all spectral components of the packet, so the carrier group index is representative of the packet as a whole.

To analyse the mapping, it is useful to introduce

\begin{equation}
    L(\tilde n_{1})
    =
    \frac{(\xi - \tilde n_{1})(n_{1} - \xi)}{n_{1}(\tilde n_{1} - n_{1})},
    \label{eq:S_L_def}
\end{equation}

\noindent which is the common value of the two fractions in Eq. \ref{eq:S_baseband-normal}. Solving for \(\tilde n_{2}\) in terms of \(L\) gives

\begin{equation}
    \tilde n_{2}(L)
    =
    \frac{L n_{2}^{2} - \xi^{2} + \xi n_{2}}{L n_{2} - \xi + n_{2}}.
    \label{eq:S_n2_of_L}
\end{equation}

\noindent The composite map \(\tilde n_{1} \mapsto \tilde n_{2}\) is then specified by \(\tilde n_{2}(\tilde n_{1}) = \tilde n_{2}[L(\tilde n_{1})]\).

\noindent We define the space-time optical push broom regime as one in which the transmitted group index \(\tilde n_{2}\) depends only weakly on the incident group index \(\tilde n_{1}\) over a finite design interval \(\mathcal{I}\),

\begin{equation}
    \left\lvert \frac{\mathrm d \tilde n_{2}}{\mathrm d \tilde n_{1}} \right\rvert \ll 1
    \quad\text{for all}\quad
    \tilde n_{1}\in\mathcal{I}.
    \label{eq:S_pb_condition}
\end{equation}

\noindent Differentiating Eqs. \ref{eq:S_L_def} and \ref{eq:S_n2_of_L} and using the chain rule yields

\begin{equation}
    \frac{\mathrm d \tilde n_{2}}{\mathrm d L}
    =
    \frac{n_{2}(n_{2}-\xi)^{2}}{[L n_{2} - \xi + n_{2}]^{2}},
    \qquad
    \frac{\mathrm d L}{\mathrm d \tilde n_{1}}
    =
    \frac{(n_{1}-\xi)^{2}}{n_{1}(\tilde n_{1}-n_{1})^{2}}.
\end{equation}

\noindent Eliminating \(L\) in favour of \(\tilde n_{2}\) through Eq. \ref{eq:S_n2_of_L} gives a symmetric expression for the slope of the mapping,

\begin{equation}
    \frac{\mathrm d \tilde n_{2}}{\mathrm d \tilde n_{1}}
    =
    \frac{n_{2}}{n_{1}}
    \frac{(n_{1}-\xi)^{2}}{(n_{2}-\xi)^{2}}
    \frac{(\tilde n_{2}-n_{2})^{2}}{(\tilde n_{1}-n_{1})^{2}}.
    \label{eq:S_slope_final}
\end{equation}

\noindent The global scale of the slope is set by \((n_{1}-\xi)^{2}/(n_{2}-\xi)^{2}\), while its variation across the design window is controlled by the distances of \(\tilde n_{1}\) and \(\tilde n_{2}\) from the luminal values \(n_{1}\) and \(n_{2}\).

\subsection{Quantitative conditions for efficient push broom}

\noindent Equation \ref{eq:S_slope_final} shows that the slope vanishes as \(\tilde n_{2}\to n_{2}\), provided \(\tilde n_{1}\) remains separated from \(n_{1}\). A strictly luminal transmitted packet, however, lies outside the validity range of the baseband approximation and is associated with unbounded spectral compression. A useful push broom regime therefore requires \(\tilde n_{2}\) to be close to, but not exactly at, \(n_{2}\), so that Eq. \ref{eq:S_slope_final} yields a small but finite slope.

The prefactor \((n_{1}-\xi)^{2}/(n_{2}-\xi)^{2}\) offers additional control. When \(\xi\) is close to \(n_{1}\), the factor \((n_{1}-\xi)^{2}\) suppresses the slope as long as \(\xi\) is not simultaneously close to \(n_{2}\). When \(\xi\) lies between \(n_{1}\) and \(n_{2}\), both \((n_{1}-\xi)\) and \((n_{2}-\xi)\) are of order unity and compression is mainly controlled by the ratio \((\tilde n_{2}-n_{2})^{2}/(\tilde n_{1}-n_{1})^{2}\). In this intermediate regime the mapping tends to compress large excursions of \(\tilde n_{1}\) into modest variations of \(\tilde n_{2}\).

Further insight comes from the limit of large incident group index magnitude. From Eq. \ref{eq:S_L_def},

\begin{equation}
    L_{\infty}
    =
    \lim_{\lvert \tilde n_{1}\rvert\to\infty} L(\tilde n_{1})
    =
    \frac{\xi - n_{1}}{n_{1}},
\end{equation}

\noindent so Eq. \ref{eq:S_n2_of_L} implies that

\begin{equation}
    \tilde n_{2}^{(\infty)}
    =
    \tilde n_{2}(L_{\infty})
    =
    \frac{\xi^{2} n_{1} - \xi n_{1} n_{2} - \xi n_{2}^{2} + n_{1} n_{2}^{2}}{\xi (n_{1}-n_{2})}
    \label{eq:S_n2_infty}
\end{equation}

\noindent is finite and independent of \(\tilde n_{1}\). Thus very large variations of the incident group index are mapped to a finite accumulation value \(\tilde n_{2}^{(\infty)}\). By choosing \(\xi\) such that \(\tilde n_{2}^{(\infty)}\) lies close to, but not exactly at, \(n_{2}\), one can realise a broad design interval where \(\lvert \mathrm d \tilde n_{2}/\mathrm d \tilde n_{1}\rvert\) is small and the transmitted spectrum remains well behaved. This provides a controlled push broom regime without the spectral divergence associated with exact luminal operation.

A complementary design strategy uses interface velocities near the phase velocity in medium-1. Writing \(\xi = n_{1} + \delta\xi\) with \(\lvert\delta\xi\rvert \ll n_{1}\), the prefactor in Eq. \ref{eq:S_slope_final} scales as \(\delta\xi^{2}\), so the slope can be made small while keeping \(\tilde n_{1}\) and \(\tilde n_{2}\) away from the luminal points. In this configuration \(\tilde n_{1}\) can span a wide range of group velocities, yet Eq. \ref{eq:S_slope_final} ensures that they are compressed into a narrow interval of \(\tilde n_{2}\).

\subsection{Co-moving STWPs and fixed point structure}

The refraction law also singles out a special class of baseband STWPs that move with the interface. Setting \(\tilde n_{1}=\xi\) in Eq. \ref{eq:S_L_def} gives

\begin{equation}
    L(\tilde n_{1}=\xi) = 0,
\end{equation}

\noindent and inserting this into Eq. \ref{eq:S_n2_of_L} yields

\begin{equation}
    \tilde n_{2}(L=0)
    =
    \frac{-\xi^{2} + \xi n_{2}}{-\xi + n_{2}}
    =
    \xi.
\end{equation}

\noindent Thus \(\tilde n_{1}=\tilde n_{2}=\xi\) is a fixed point of the mapping, and an STWP that co-moves with the interface retains essentially the same group index after refraction.

A linear expansion around this fixed point shows that the slope of the mapping is of order unity rather than strongly suppressed, so nearby STWPs do not experience significant push broom compression. Instead, their group indices are transferred with only modest distortion. A detailed study of this co-moving sector, including higher order corrections and finite bandwidth effects, lies beyond the scope of the present work but may be useful for applications that require selective isolation of co-moving packets rather than velocity compression.

\section{Velocity spectral optical fission details}

\subsection{Quadratic form of the X--wave refraction law}

For X--wave STWPs at normal incidence, the space-time refraction law derived in the main text can be written in terms of the group indices \(\tilde n_{1}\) and \(\tilde n_{2}\), the refractive indices \(n_{1}\) and \(n_{2}\), and the interface parameter \(\xi=c/v_{\mathrm m}\) as

\begin{equation}
    \frac{\tilde n_{1} - \xi}{\sqrt{n_{1}^{2} - \tilde n_{1}^{2}}}
    =
    \frac{\tilde n_{2} - \xi}{\sqrt{n_{2}^{2} - \tilde n_{2}^{2}}}.
\end{equation}

\noindent with the existence condition \(n_{j}^{2}-\tilde n_{j}^{2}>0\). Introducing the invariant

\begin{equation}
    K
    =
    \frac{\tilde n_{1} - \xi}{\sqrt{n_{1}^{2} - \tilde n_{1}^{2}}},
\end{equation}

\noindent the transmitted branches in medium-2 satisfy

\begin{equation}
    \frac{\tilde n_{2} - \xi}{\sqrt{n_{2}^{2} - \tilde n_{2}^{2}}}
    =
    K.
\end{equation}

\noindent Squaring and rearranging gives a quadratic equation for \(\tilde n_{2}\),

\begin{equation}
    (1+K^{2})\,\tilde n_{2}^{2}
    - 2\xi\,\tilde n_{2}
    + (\xi^{2} - K^{2} n_{2}^{2}) = 0,
\end{equation}

\noindent with candidate solutions

\begin{equation}
    \tilde n_{2}^{(\pm)}
    =
    \frac{\xi \pm K \sqrt{(1+K^{2})n_{2}^{2} - \xi^{2}}}{1+K^{2}}.
\end{equation}

\noindent The discriminant

\begin{equation}
    \Delta
    =
    4 K^{2}\bigl[(1+K^{2}) n_{2}^{2} - \xi^{2}\bigr]
\end{equation}

\noindent controls the number of real roots. A necessary condition for double refraction is \(\Delta>0\), which requires

\begin{equation}
    (1+K^{2}) n_{2}^{2} - \xi^{2} > 0.
\end{equation}

\noindent Physical X--wave branches additionally obey \(\lvert\tilde n_{2}^{(\pm)}\rvert < n_{2}\) and must satisfy the unsquared refraction law with a consistent sign choice.

\subsection{Domain of double refraction}

For given \((n_{1},n_{2},\xi)\) and incident group index \(\tilde n_{1}\), the parameter \(K\) is fixed by
\begin{equation}
K
=
\frac{\tilde n_{1} - \xi}{\sqrt{n_{1}^{2} - \tilde n_{1}^{2}}}.
\end{equation}

\noindent The double refraction window corresponds to incident indices for which

1. \(n_{1}^{2} - \tilde n_{1}^{2} > 0\) (existence of the incident X--wave),

2. \((1+K^{2}) n_{2}^{2} - \xi^{2} > 0\) (positive discriminant),

3. both \(\tilde n_{2}^{(\pm)}\) satisfy \(\lvert\tilde n_{2}^{(\pm)}\rvert < n_{2}\),

4. the signs in the square roots are chosen so that the original unsquared relation between the two sides of the refraction law holds.

\noindent When these conditions are satisfied, one incident X--wave with group index \(\tilde n_{1}\) produces two transmitted X--waves with group indices \(\tilde n_{2}^{(+)}\) and \(\tilde n_{2}^{(-)}\). Their group velocities are

\begin{equation}
v_{g2}^{(\pm)} = \frac{c}{\tilde n_{2}^{(\pm)}}.
\end{equation}

For the parameters used in Fig. 3 of the main text, \(n_{1}=1.5\), \(n_{2}=1.0\), and \(v_{\mathrm m}=0.5 c\) so that \(\xi=2\), these inequalities define a narrow interval of incident tilt angles \(\theta_{1}\) (equivalently \(\tilde n_{1}=\cot\theta_{1}\)) where both branches \(\tilde n_{2}^{(\pm)}\) are physical. This interval corresponds to the green band in Fig. 3(a) in the main article, and the two branches are labelled STWP-1 and STWP-2.

\subsection{Real space interpretation}

The quadratic structure of the X--wave refraction law has a simple geometric origin. In \((k_{x},k_{z},\omega)\) space, the moving interface selects a space-time refraction plane specified by the Doppler invariant \(\omega - v_{\mathrm m} k_{z}\). The incident X--wave spectrum is the intersection of this plane with the light cone of medium-1 and an X--wave plane through the origin, which gives a straight conic section. The same refraction plane intersects the light cone of medium-2 along either one or two curves in the physical octant. When there are two such curves and both satisfy the X--wave existence condition, the refraction law becomes double valued and produces two transmitted X--waves with different spectral tilts.

In the time domain this manifests as a splitting of the incident world line into two transmitted world lines with different slopes. Writing the group velocities as
\begin{equation}
    v_{g1} = \frac{c}{\tilde n_{1}},
    \qquad
    v_{g2}^{(\pm)} = \frac{c}{\tilde n_{2}^{(\pm)}},
\end{equation}

\noindent the axial trajectories of the packet centres at \(x=0\) can be approximated by

\begin{equation}
    z_{1}(t) \simeq v_{g1} t,
    \qquad
    z_{2}^{(\pm)}(t) \simeq z_{0} + v_{g2}^{(\pm)} (t - t_{0}),
\end{equation}

\noindent where \(t_{0}\) and \(z_{0}\) mark the interaction with the interface. In the example of Fig. 3, one transmitted branch has \(v_{g2}^{(+)} > v_{g1}\) and moves ahead of the interface, while the other has \(v_{g2}^{(-)} < v_{g1}\) and falls behind. The intensity snapshots in the main text confirm that the spatial profiles of both branches remain X-shaped, so each transmitted packet is propagation invariant in its own group velocity frame even though they originate from a single incident X--wave.

\end{document}